\documentclass[aps,prev,preprintnumbers,floatfix,nofootinbib]{revtex4-1}
\pdfoutput=1

\usepackage{mathrsfs, amsmath, amsthm, amssymb, color, epstopdf, verbatim, hyperref, enumerate, graphicx, hyperref,ulem}

\usepackage{titlesec}

\hypersetup{
	colorlinks,
	citecolor=blue,
	filecolor=black,
	linkcolor=blue,
	urlcolor=black
}

\usepackage{xcolor}
\definecolor{MSBlue}{rgb}{.204,.353,.541}
\definecolor{MSLightBlue}{rgb}{.310,.506,.741}


\titleformat*{\section}{\Large\bfseries}
\titleformat*{\subsection}{\Large\bfseries}
\titleformat*{\subsubsection}{\Large\bfseries}

\begin{document}

\title{A detailed exploration of the EDGES 21 cm absorption anomaly and axion-induced cooling}
\author{Chuang Li$\,{}^{a}$}
\email{lichuang\_physics@163.com}
\author{Nick Houston$\,{}^b$}
\email{nhouston@bjut.edu.cn, corresponding author}
\author{Tianjun Li$\,{}^{c,d}$}
\email{tli@itp.ac.cn}
\author{Qiaoli Yang$\,{}^{e}$}
\email{qiaoli\_yang@hotmail.com}
\author{Xin Zhang$\,{}^{f,g}$}
\email{zhangxin@nao.cas.cn}
\affiliation{
	${}^a$ College of Mechanical and Electrical Engineering, Wuyi University, Nanping 354300, China\\
	${}^b$ Faculty of Science, Beijing University of Technology, Beijing 100124, China\\
	${}^c$ CAS Key Laboratory of Theoretical Physics, Institute of Theoretical Physics, Chinese Academy of Sciences, Beijing 100190, China\\
	${}^d$ School of Physical Sciences, University of Chinese Academy of Sciences, No. 19A Yuquan Road, Beijing 100049, China\\
	${}^e$ Siyuan Laboratory and Physics Department, Jinan University, Guangzhou 510632, China\\
	${}^f$ Key Laboratory of Computational Astrophysics, National Astronomical Observatories, Chinese Academy of Sciences, 20A Datun Road, Chaoyang District, Beijing 100012, China\\
	${}^g$ School of Astronomy and Space Science, University of Chinese Academy of Sciences, No. 19A Yuquan Road, Beijing 100049, China
}

\begin{abstract}
The EDGES collaboration's observation of an anomalously strong 21 cm absorption feature around the cosmic dawn era has energised the cosmological community by suggesting a novel signature of dark matter in the cooling of cosmic hydrogen.
In a recent letter we have argued that by virtue of the ability to mediate cooling processes whilst in the condensed phase, a small amount of axion dark matter can explain these observations within the context of standard models of axions and axion-like particles.
These axions and axion-like particles (ALPs) can thermalize through gravitational self-interactions and so eventually form a Bose-Einstein condensate (BEC), whereupon large-scale long-range correlation can produce experimentally observable signals such as these.
In this context the EDGES best-fit result favours an axion-like-particle mass in the (6, 400) meV range.
Future experiments and galaxy surveys, particularly the International Axion Observatory (IAXO) and EUCLID, should have the capability to directly test this scenario.
In this paper, we will explore this mechanism in detail and give more thorough computational details of certain key points.

\end{abstract}

\maketitle
\tableofcontents

\section{Introduction}

In the standard cosmology, our present Universe arose from a hot soup of radiation and matter left over after the big bang.
After the first 300,000 years elapsed, sufficient cooling allowed free electrons and protons to combine to form neutral hydrogen atoms and thus begin the cosmological dark ages, an epoch in which there were no luminous sources.
Over time, local overdensities collapsed to form the first generations of stars and galaxies, resulting in the so-called cosmic dawn.
As time went on this process continued, rendering the universe fully reionized, and ultimately resulting in today's almost transparent intergalactic medium dotted with galaxies, quasars and galaxy clusters.

Although our understanding of cosmology has advanced significantly over the last decades, many aspects of this cosmic dawn are still unexplored.
Meanwhile, in the standard $\Lambda$CDM cosmology, this epoch is relatively predictable and easily understood.
As such, it can serve as a ideal probe of physics beyond the $\Lambda$CDM paradigm.

One key observable is related to the absorption or emission spectrum of cosmic microwave background radiation at the cosmic dawn, arising from the neutral hydrogen present absorbing wavelengths close to atomic transitions, and thereby imprinting a characteristic spectral distortion in the vicinity of 21 cm, by virtue of the singlet/triplet spin flip transition.
This feature redshifts to about 80 MHz today, and has been observed by the Experiment to Detect the Global Epoch of reionization Signature (EDGES) Collaboration~\cite{Bowman:2018yin}, who claim that the effective 21 cm brightness temperature is: 
\begin{equation}
T_{21} \simeq -0.52^{+0.18}_{-0.42}\,\mathrm{K} 
\end{equation}
where the uncertainties quoted are at 99\% confidence level.

The amplitude of this signal, fully described by subsequent Eq.\eqref{T21_Full}, is given here in simple terms as
\begin{equation}
	T_{21} \simeq 35 \mathrm{mK}\left(1-\frac{T_\gamma}{T_s}\right)\sqrt{\frac{1+z}{18}}
	\label{T21}
\end{equation}
where $T_s$ is the singlet/triplet spin temperature(see Eq.\eqref{Ts_Def} for the exact definition) of the hydrogen gas present at that time, and $T_\gamma$ is the CMB temperature.
Once stellar emission of UV radiation begins, perhaps around $z \sim 20$, we expect that $T_\gamma >> T_s \gtrsim T_{\mathrm{gas}}$, due to the decoupling of hydrogen gas and the CMB at earlier times, and the coupling of the spin temperature to the kinetic gas temperature around this period. In the standard $\Lambda$CDM scenario, $T_\gamma|_{z \sim 17} \simeq 49$ K and $T_{\mathrm{gas}}|_{z \sim 17} \simeq 6.8$ K, so by Eq.\eqref{T21}, we have 
\begin{equation}
T_{21} \gtrsim -0.21 K
\end{equation}
The significance of the EDGES deviation from the $\Lambda$CDM expectation is estimated to be $3.8\,\sigma$, which represents an anomalously strong 21cm absorption feature from $z\in (20,15)$, corresponding to the era of early star formation.

Explanations of the EDGES result proposed in the literature can be roughly classified into four types.
Firstly, new mechanisms for cooling the gaseous medium, such as in Refs.~\cite{Barkana:2018lgd, Munoz:2018pzp, Berlin:2018sjs, Barkana:2018qrx, Li:2018kzs, Lambiase:2018lhs, Houston:2018vrf, Sikivie:2018tml, Slatyer:2018aqg, Falkowski:2018qdj} by some kind of dark matter or ~\cite{Costa:2018aoy, Hill:2018lfx} by dark energy.
Secondly, proposals adding an extra soft photon component to promote the CMB temperature, such as in Refs.~\cite{Feng:2018rje, Ewall-Wice:2018bzf, Fraser:2018acy, Pospelov:2018kdh, Lawson:2018qkc, Moroi:2018vci, AristizabalSierra:2018emu}. 
Thirdly, modifications of cosmological evolution process, such as in Refs.~\cite{Wang:2018azy, Xiao:2018jyl}.
Lastly, improvements to the measurement and treatment of the foregrounds to relax the anomaly, such as in Refs.~\cite{Hills:2018vyr, Xu:2018shq}.

We are interested in the first of these approaches, but the interaction cross section required to achieve this is prohibitive for most models of dark matter \cite{Barkana:2018lgd}.
To be consistent with other experimental and observational constraints, models capable of explaining the EDGES observation require millicharged dark matter comprising just $0.3-2$\% of the total dark matter abundance, with masses and millicharges in the $(10, 80)$ MeV and $(10^{-4},10^{-6})$ ranges, respectively~\cite{Munoz:2018pzp, Berlin:2018sjs, Barkana:2018qrx}.

As such, in the following we will propose a more natural dark-matter theoretic approach, which is based on the speculated ability of axion dark matter to form a Bose-Einstein Condensate \cite{Sikivie:2009qn,Erken:2011dz}.
In many respects, this condensed state behaves as ordinary CDM (Cold Dark matter), but it has a particularly interesting ability to induce transitions between momentum states of coupled particle species, and hence is naturally equipped with a cooling effect.
This mechanism was originally invoked in Ref.~\cite{Erken:2011vv} to lower the photon temperature in the era of Big Bang Nucleosynthesis (BBN), albeit with a different motivation to our own \footnote{Concretely, the authors of~\cite{Erken:2011vv} wanted to ease the discrepancy between the observed and predicted primordial ${}^7$Li abundance by adjusting the baryon-to-photon ratio.}.

By consistently adjusting the parameter range, we can analogously lower the hydrogen temperature prior to the cosmic dawn to explain the EDGES observations in the context of axion and axion-like-particle (ALP) models.
We fortunately find that, to be close to the existing experimental limits, the implied parameter range could be tested at the next generation of axion experiments and large scale surveys, particularly IAXO\cite{Armengaud:2014gea} and EUCLID\cite{Laureijs:2011gra}.

This paper is intended to serve as an extended edition of a previous letter ``Natural explanation for 21 cm absorption signals via axion-induced cooling''~\cite{Houston:2018vrf}.
As such the content will be similar, but we will provide more background and detailed computational steps.

We emphasize for clarity that, although several other axion-related explanations~\cite{Lambiase:2018lhs,Lawson:2018qkc,Moroi:2018vci} have been proposed, our approach differs in many essential respects from them. 
We also note Ref.~\cite{Sikivie:2018tml}, which appeared shortly after Ref.~\cite{Houston:2018vrf} deals with the same scenario of axion BEC-induced cooling and 21 cm cosmology, but with different emphasis.

The outline of this paper is as follows.
In Section II, we give a concise introduction of the astrophysics background, including 21 cm cosmology and the EDGES experiment.
In Section III, we give an overview of the relevant aspects of axion physics, covering the basic motivation of the axion proposal, the axion dark matter scenario, and the Bose-Einstein condensation of axion dark matter.
In Section IV, the hydrogen cooling mechanism is explored in detail, based on the axion BEC, including concrete cooling rate formulae.
In Section V, we demonstrate the resulting parameter space constraints, and discuss the possible experiments and observations with the potential to confirm this model.
Section VI provides a summary.

\section{Astrophysics background}

\subsection{The 21 cm signal}

The so-called 21 cm signal/line is associated to the hyperfine splitting between the spin singlet and triplet states of the electron and proton in a hydrogen atom.
Hydrogen is ubiquitous in the Universe, amounting to $\sim 75\%$ of the gas mass present in the intergalactic medium (IGM). As such, this observable provides a convenient tracer of the properties of the first billion years of our Universe.
The 21 cm line from gas during this time redshifts to 30-200 MHz today, making it a prime target for the new generation of radio interferometers currently being built.
The related research is an active area of astrophysics and cosmology, often referred to simply as `21 cm cosmology'~\cite{Madau:1996cs, Pritchard:2011xb}.

There are two kinds of 21 cm signal, full 3D and global (sky-averaged), respectively.
For the full 3D signal, observations of the 21 cm line constrain the properties of the intergalactic medium and the cumulative impact of light from all galaxies. In combination with other direct observations of the sources, they provide a powerful tool for learning about the first stars and galaxies. By observing the surrounding ionization bubbles, they can also provide useful information about active galactic nuclei (AGN), such as quasars.
Because part of the signal couples with the density field, which can give some information about the initial conditions of cosmic inflation and neutrino masses in the form of the power spectrum, it could also allow precise measurements of cosmological parameters, thereby illuminating fundamental physics.

Besides the above physical significance of the full 3D 21 cm signal, Ref.~\cite{Mirocha:2013gvq,Mirocha:2015jra,Mirocha:2017xxz} has also shown that the global (sky-averaged) 21 cm observations can be used to constrain for example the Lyman-$\alpha$ background intensity and heat deposition, the growth rate of dark matter halos, and to provide unique signatures of Population III stars.

\subsection{Detection mechanism}

The detectability of the 21 cm signal relies on the spin temperature, an effective temperature that describes the relative abundances of the ground and excited states of the hyperfine splitting of the hydrogen atom, defined by
\begin{equation}
	n_1/n_0=(g_1/g_0)\exp(-T_\star/T_s),
	\label{Ts_Def}
\end{equation}
where $n_i$ are the number densities of hydrogen atoms in the two hyperfine levels, subscript 0 and 1 for the $1S$ singlet and $1S$ triplet levels, respectively, $g_i$ is the statistical degeneracy factors of the two levels, with $(g_1/g_0) = 3$, and $T_\star \equiv hc/k\lambda_{21\rm{cm}} = 0.068\,\rm{K}$.
This signal can only be observed when the spin temperature deviates from a given background.

In the early Universe when the gas density is high, collisions between different particles may induce spin-flips in a hydrogen atom and dominate the spin temperature coupling\footnote{There exist three main channels: collisions between two hydrogen atoms, collisions between a hydrogen atom and an electron, and similarly, collisions between a hydrogen atom and a proton.}.
Then the spin temperature can be identified with the background gas temperature, which is the same as the CMB temperature.

For most of the redshifts $z \le 200$, gas collisional coupling of the 21 cm line is inefficient, and absorption/emission of 21 cm photons to and from the radio background, primarily the CMB, becomes the most important element affecting the spin temperature, so the spin temperature is close to the CMB temperature. 

However, when the first generations of stars begin to form (around $z \sim 20$), Ly$\alpha$ photons offer another coupling channel via the so-called Wouthuysen-Field effect \cite{wouth1952,field1958}.
The main idea is as follows: suppose that hydrogen is initially in the hyperfine singlet state, so that absorption of a Ly$\alpha$ photon will excite the atom into higher energy level, such as either of the central 2P hyperfine states.
From here, the transition to a lower energy level can place the atom in either of the two ground state hyperfine levels (i.e., the spin singlet and triplet states).
If the final state is triplet then a spin-flip has occurred. 
Hence, Ly$\alpha$ photons can induce spin-flips via an intermediate excited state.

As a whole, the spin temperature is determined by three processes:  {\it (i)} collisions with other hydrogen atoms, electrons and protons; {\it (ii)} absorption/emission of 21 cm photons to and from the CMB; {\it (iii)} scattering of Ly$\alpha$ photons.
The resulting spin temperature is then set by the equilibrium balance of these effects \footnote{The equilibrium condition is an excellent approximation since the rate of these processes is much faster than the de-excitation time of the 21 cm line.}, formulated as
\begin{equation}
	T_s^{-1}=\frac{T_\gamma^{-1}+x_\alpha T_\alpha^{-1}+x_c T_K^{-1}}{1+x_\alpha+x_c}
	\label{Ts_Full}
\end{equation}
where $T_\gamma$ is the temperature of CMB, $T_\alpha$ is the color temperature of the Ly$\alpha$ radiation, $T_K$ is the gas kinetic temperature, $T_\alpha$ is closely coupled to $T_K$ from the repeated recoil and scattering, and $x_\alpha, x_c$ are the coupling coefficients due to scattering of Ly$\alpha$ photons and atomic collisions, respectively \cite{field1958}.
Obviously, the spin temperature will be very close to $T_\gamma$ when $x_\alpha + x_c \ll 1$ and approaches the gas temperature when $ x_\alpha + x_c \gg 1$.

Roughly speaking, shortly after the decoupling of the CMB and hydrogen gas at $z \sim 200$, $T_s$ is between $T_\gamma$ and $T_H$; as the Universe expands, diluting the gas, $T_s$ moves towards $T_\gamma$.
Around $z \sim 20$, stars begin the emission of Ly$\alpha$ and X-ray photons, and $T_s$ couples to the hydrogen gas temperature due to the Wouthuysen-Field effect.
So we expect that $T_\gamma \gg T_s \gtrsim T_{\mathrm{gas}}$ at $z \sim 20$.

To be used conventionally with observations, Ref.~\cite{Madau:1996cs} also defines an effective 21 cm brightness temperature
\begin{equation}
	T_{21} = 26.8\, x_{\rm HI}\, \frac{\rho_{\rm g}}{\bar{\rho}_{\rm g}} \left( \frac{\Omega_{\rm b}
	h} {0.0327} \right) \left(\frac{\Omega_{\rm m}}{0.307}\right)^{-1/2} \left( \frac{1+z} {10} \right)^{1/2} \left(\frac{T_s-T_\gamma} {T_s} \right)\, \mathrm{mK}
	\label{T21_Full}
\end{equation}
where $x_{\rm HI}$ is the mean mass fraction of hydrogen that is neutral (i.e. not ionized), $\rho_{\rm g}$ is the gas density and $\bar{\rho}_{\rm g}$ its cosmic mean value, $\Omega_{\rm m}$ and $\Omega_{\rm b}$ are the cosmic mean densities of matter and of baryons, respectively, in units of the critical density, $h$ is the Hubble parameter in units of $100\, \mbox{ km s}^{-1}\, \mbox{Mpc}^{-1}$, $z$ is the redshift (corresponding to an observed wavelength of 21$\times (1+z)$~cm and an observed frequency of $1420/(1+z)$~MHz), $T_\gamma=2.725\times(1+z)$ is the CMB temperature at $z$, and $T_s$ is the spin temperature of hydrogen at $z$ as shown in Eq.\eqref{Ts_Full}.

\subsection{Detection instruments and the EDGES observation}

As we have shown in the previous section, detecting the 21 cm signal from the cosmic dawn should enable key insights into the nature of the first stellar objects and their substantial influence on galaxy formation, and the later processes which lead to the complex structures we see in the Universe today.
The instruments to do so are roughly classified in two ways: full 3D signal detection instruments, and global signal detection instruments.
The global signal can be viewed as a zeroth order approximation to the full 3D signal, as it is averaged over large angular scales.

Because of the spatial variation in the different radiation fields and properties of the IGM, the full 3D signal will be highly inhomogeneous.
There are many full 3D signal detection instruments, such as the Giant Meterwave Radio Telescope (GMRT)\cite{1990IJRSP..19..493S, 1991ASPC...19..376S}, PrimevAl Structure Telescope (PAST or 21CMA)\cite{Pen:2004de}, Murchison Widefield Array (MWA)\cite{Lidz:2007az, Lonsdale:2009cb}, the LOw Frequency ARray (LOFAR)\cite{vanHaarlem:2013dsa}, and the Precision Array to Probe the Epoch of Reionization (PAPER)\cite{2005AAS...207.3301B}, which aim to detect the fluctuations of the redshifted 21 cm radio background induced by variations in the neutral hydrogen density.
Next generation instruments, such as SKA \cite{Braun:1995da, Carilli:2004nx, Kanekar:2004nk, 2004SPIE.5489...62S, Feretti:2017ade, Bull:2018lat}, will be able to make more detailed observation of the ionized regions during reionization, and probe more of the properties of cosmic hydrogen.

The global 21 cm signal is averaged over the sky without high angular resolution, so it can be detected as an absolute frequency-dependent temperature via a single dipole antenna.
The pioneering experiments aiming to do so are the Sonda Cosmológica de las Islas para la Detección de Hidrógeno Neutro (SCI-HI)\cite{Voytek:2013nua}, the Large-Aperture Experiment to Detect the Dark Ages (LEDA)\cite{Greenhill:2012mn},  the Shaped Antenna measurement of the background Radio Spectrum 2 (SARAS 2)\cite{Singh:2017gtp}, the COsmological Reionization Experiment (CORE)\cite{CORE:All}, the Probing Radio Intensity at high z from Marion (PRIZM)\cite{2018arXiv180609531P}, and the Experiment to Detect the Global EoR Signature (EDGES)\cite{Bowman:2018yin, Barkana:2018lgd}.

EDGES is a collaboration between Arizona State University and the MIT Haystack Observatory, funded by the National Science Foundation (NSF).
The project’s goal is to detect the radio signatures of hydrogen from the cosmic period known as the Epoch of Reionization (EoR), soon after the formation of the first stars and galaxies.
Concretely, EDGES aims to probe the radiative properties of the first stars and compact objects via the profile of the observed 21 cm absorption or emission spectrum features in the radio background.

Although the detection principle is simple, a practical measurement is complicated by the need for nontrivial subtraction of galactic foregrounds, which are orders of magnitude larger than the expected signal.
On the assumption of spectral smoothness of the foreground signals, in contrast to the specific spectral structure of the 21 cm absorption profile, foregrounds can be predicted by, for example, fitting a low order polynomial, which once subtracted leaves the desired 21 cm signal in the residuals \cite{2016MNRAS.455.3890M}. 
In foreground estimation an additional complication is the Earth’s ionosphere, which affects the propagation of radio waves via absorption of incoming radiation, and direct thermal emission from electrons in the ionosphere. 
This contribution can however also be modeled with a good fit to the data \cite{Rogers-2015}.
To account for possible mixing with the foregrounds, precise calibration of the instrumental frequency response is also required, which is a significant milestone completed by the EDGES collaboration \cite{Monsalve:2016xbk}.
Note: While the present paper was being completed, a new controversy about foreground model consistency appeared\cite{Foreground_Model_Controversy}, however the response of EDGES collaboration is substantive\cite{Foreground_Model_Defense}.
With global 21 cm signal model\cite{Pritchard:2011xb}, EDGES can then perform least-squares and/or Markov Chain Monte Carlo (MCMC) analyses to get best-fits and reasonable confidence intervals, and consequently derive astrophysical information from observations through parameter estimation in these calibrated and integrated model fits \cite{2011MNRAS.411..955M, 2017MNRAS.472.1915C}.

\section{Axion physics}

In this section we will give a brief overview of the relevant aspects of axion physics and cosmology, further details are available in Ref.~\cite{Yang:2015qka, Kim:2017yqo, Marsh:2015xka}.

\subsection{Basic properties}

The axion is a hypothetical elementary particle originally postulated in 1977 \cite{Peccei:1977hh, Weinberg:1977ma, Wilczek:1977pj} to resolve the strong CP problem in quantum chromodynamics (QCD).
The existence of axion like particles (ALPs) are more generally predictions of many high energy physics models, including string theory in particular.
In the following, the QCD axion is contrasted with ALPs, while we use the term `axion' flexibly to include both the QCD axion and ALPs.

~\\
\textbf{The QCD axion}

The QCD axion primarily provides a excellent solution to the strong CP problem, and thus an attractive target for particle physics searches beyond the Standard Model\cite{Peccei:1977hh, Weinberg:1977ma, Wilczek:1977pj, Kim:1979if, Shifman:1979if, Dine:1981rt, Zhitnitsky:1980tq}.
The basic idea is concisely described as follows.
The QCD vacuum is non-trivially dependent on a parameter $\theta\in[0,~2\pi]$, which results an effective CP odd term to the QCD Lagrangian,
\begin{equation}
	\mathcal{L}_{\theta} = \frac{\theta}{32\pi^2} G^{a}_{\mu\nu}\tilde{G}^{\mu\nu a}
	\label{eqn:qcd_topological}
\end{equation}
where $G^{a}_{\mu\nu}$ is the gluon field strength tensor, and $\tilde{G}^{\mu\nu a}=\epsilon^{\mu\nu\alpha\beta}G^{a}_{\alpha\beta}/2$ is its dual.
This $\theta$ term can be changed by adding additional phases to the quark mass matrix via $q_i\to e^{i\alpha_i\gamma_5/2}q_i$, and so the physical parameter that determines the CP violation in QCD is $\tilde\theta$, 
\begin{equation}
	\tilde \theta  = \theta-\sum\alpha_i
\end{equation}
or	
\begin{equation}
	\tilde \theta = {\theta} + \text{arg det}M_u M_d
\end{equation}
where $M_u$, $M_d$ are the quark mass matrices.

The term $\mathcal{L}_{\tilde\theta}$ allows P and T or CP violation, but since CP is conserved well in QCD processes, we infer that$ |\tilde \theta |\lesssim 10^{-10} $ ~\cite{Crewther:1979pi,Baker:2006ts}. 
This fine tuning constitutes the strong CP problem.

The first solution is proposed by Peccei \& Quinn~\cite{Peccei:1977hh}, which introduced an $U(1)$ symmetry (PQ symmetry), and a scalar (the PQ scalar) which dynamically leads to $\tilde\theta\simeq0$ due to the effects of a potential for $\tilde\theta$ arising from QCD instanton effects. 
Later this model was complemented by Wilczek \& Weinberg by implementing the same mechanism (the PQ mechanism) in the realistic Weinberg-Salam model~\cite{Wilczek:1977pj, Weinberg:1977ma}, where they realized that the spontaneous breaking of PQ symmetry results in a pseudo-Goldstone boson, coined the axion by Wilczek.
The above Peccei-Quinn-Wilczek-Weinberg (PQWW) model was excluded by experiment quickly, but the PQ mechanism was subsequently implemented by other models, the most important of which are the following two.

\begin{itemize}	
	\item The Kim-Shifman-Vainshtein-Zakharov (KSVZ)~\cite{Kim:1979if, Shifman:1979if} axion model, which adds the PQ scalar and a heavy quark to the Weinberg-Salam model.
	\item The Dine-Fischler-Srednicki-Zhitnitsky (DFSZ)~\cite{Dine:1981rt, Zhitnitsky:1980tq} axion model, which adds the PQ scalar and an additional Higgs field to the Weinberg-Salam model.
\end{itemize}

The PQ mechanism is also endorsed by a theorem of Vafa and Witten~\cite{Vafa:1984xg}  which claims that the vacuum energy of the QCD instanton potential is minimized at the $CP$ conserving value.

To be more specific, here we will follow the literature ~\cite{Peccei:2006as, Kim:2008hd, Dine:2000cj}, to present the PQ mechanism briefly. 
The complex PQ scalar field is expressed as:
\begin{equation}
	S =\rho \exp\left(i\frac{\phi}{f_a}\right)
\label{eqn:PQ_Scalar}
\end{equation}
where $\rho$ is the radial field, $\phi$ is the angular field, and $f_a = \left<S\right>$ is the energy scale of PQ symmetry breaking. 
After PQ symmetry breaking, $\phi$ is the massless Goldstone boson of this broken symmetry, axion.
$\rho$ is generally massive and decouples, so the Lagrangian changes to:
\begin{equation}
	L_{\rm{total}}= L_{\rm{SM}} -\frac{1}{2} \partial_{\mu}\phi\partial^{\mu}\phi + L_{\rm{int}}[\partial^{\mu} \phi/f_a; \Psi] + \left(\bar{\theta} + \mathcal{C}\frac{\phi}{f_a}\right)\frac{g^2}{32\pi^2} F_a^{\mu \nu}\tilde{F}_{ a\mu \nu}. 
\end{equation}
where $\mathcal{C}$ is known as the colour anomaly of the PQ symmetry, and is given by
\begin{equation}
\mathcal{C}\delta_{ab}=2\text{Tr }\mathcal{Q}_{\rm PQ}T_aT_b
\end{equation}
Here the trace is over all the fermions in the theory, and $T_a$ are the generators of the corresponding fermion representations. The colour anomaly also represents the number of potential vacua in the range $[0,2\pi f_a]$, is thus also known as the domain wall number, which must be integer~\cite{Srednicki:1985xd}. 

The last term is from the PQ symmetry chiral anomaly, which also leads to an induced axion potential, $V_{\rm{eff}}(\phi)$. 
The dynamics of $\phi$ send it to one of these vacua, which is the essence of the PQ mechanism,
\begin{equation}
<\frac{\partial V_{\rm{eff}}}{\partial \phi}>= -\frac{\xi}{f_a}\frac{g^2}{32\pi^2}< F_a^{\mu \nu}\tilde{F}_{ a\mu \nu}>| _{<\phi>=-\frac{f_a}{\mathcal{C}} \bar{\theta}}~~=0
\end{equation}
More concretely, 
\begin{equation}
V_{\rm{eff}} \sim 1-cos[\bar{\theta} + \mathcal{C} \frac{<\phi>}{f_a} ]
\end{equation}
So, $<\phi>=-\frac{f_a}{\mathcal{C}} \bar{\theta}$ will minimize $V_{\rm{eff}}(\phi)$.
In the following paper, unless otherwise stated the colour anomaly will be absorbed into $f_a$.
For the small $\phi$ displacements from the potential minimum, the potential can be expanded as a Taylor series.
The dominant piece is the mass term:
\begin{equation}
V_{\rm{eff}}(\phi)\approx {1 \over 2} m_a^2\phi^2 
~~ or ~~
m_a^2= <\frac{\partial^2 V_{\rm{eff}}}{\partial \phi^2}>
\label{eqn:mass_potential}
\end{equation}
Furthermore, the mass is given by:
\begin{equation}
m_{a} = \Lambda_a^2/f_a \approx 6\times 10^{-6}\text{eV}\left(\frac{10^{12}\text{ GeV}}{f_a} \right)
\label{qcd_axion_mass}
\end{equation}
where $\Lambda_a$ is a non-perturbative scale induced by QCD instantons. 
Typically the PQ symmetry breaking scale $f_a$ is much higher than $\Lambda_a$, so axion gets a small mass. 
In addition, axion self-interactions and interactions with Standard Model fields are also suppressed by powers of $f_a$. 
By virtue of the underlying shift symmetry $\phi\rightarrow \phi+\text{const}$, the axion mass is protected from perturbative quantum corrections. 
In general, the axion mass will also be temperature dependent, because of the temperature dependence of the non-perturbative effects.
All of these elements conspire to make the axion a light, weakly interacting, long-lived particle, and hence a natural dark matter candidate~\footnote{These properties also permit axions to address problems of inflation or dark energy, which are out of our present scope.}.

There are of course newer variant models with some enhancements, designed to avoid issues such as Landau poles, or astrophysical limits, etc. ~\cite{DiLuzio:2016sbl, DiLuzio:2017pfr, DiLuzio:2017ogq}, or to incorporate PQ symmetry with some kinds of supersymmetric standard model or grand unified theory~\cite{Nilles:1981py, Wise:1981ry, Boucenna:2017fna, Ernst:2018bib}, or to get Peccei-Quinn symmetries from exact discrete symmetries~\cite{Dias:2014osa}, or to unify PQ symmetry and neutrino seesaw mechanism ~\cite{Ballesteros:2016euj, Ma:2017vdv}, which overall give more flexible theoretical configurations.

~\\
\textbf{Axion-like particles}

Any breaking of anomalous global symmetries will result in some pseudo Nambu-Goldstone (pNG) bosons, if these bosons have similar properties to the QCD axion we will refer them as axion-like particles (ALPs).
There are some notable examples, such as lepton number symmetry breaking~\cite{Chikashige:1980ui, Gelmini:1980re} or family/flavor symmetry breaking~\cite{Wilczek:1982rv, Berezhiani:1990wn, Jaeckel:2013uva, Arias-Aragon:2017eww}, but the most interesting situation we should mention is that ALPs commonly exist in the string theory framework \cite{String-Book:GSW, String-Book:Polchinski, String-Book:BBS}.

As is well known, extra spacetime dimensions are introduced in string theory for self-consistency. 
These additional dimensions are typically compactified to very small size manifolds (usually of ``Calabi-Yau'' type) to be compatible with our 4D world.

In the low energy effective description of string theory, the zero modes of the antisymmetric tensors on the compactified manifolds have similar properties to QCD axions~\cite{String-Book:GSW, Svrcek:2006yi}. 
For example, they are pseudoscalar, they have axion-like couplings (to $G^{a}_{\mu\nu}\tilde{G}^{\mu\nu a}$), they are massless to all orders in perturbation theory, and they can obtain a small mass via a non-perturbative instanton potential. 
So, they are called axion-like particles. 
From the point of view of phenomenology, the key differences between the QCD axion and axion-like particles are the coupling strengths to the standard model particles.

\subsection{The axion as dark matter}

Dark matter (DM) is a critical element of the modern standard model of cosmology.
If the QCD axion exists and the decay constant $f_a$ in Eq.\eqref{qcd_axion_mass} is large, it should be extremely weakly interacting and stable, and thus an excellent DM candidate \cite{AxionDM-Preskill:1982cy, AxionDM-Abbott:1982af, AxionDM-Dine:1982ah, AxionDM-Kim:1986ax, AxionDM-Sikivie:2006ni}.
Whilst the literature on this topic is abundant, we will in the following focus purely on the elements which are of relevance to our later discussion.

\subsubsection{The axion field in the early Universe}

For the axion field in the early Universe, there are several entangled factors of relevance. 
For a clear description, we briefly delineate these into three parts as follows.

~\\
\textbf{Productions of axions}

As mentioned in the previous section, the evolution of axions in the early Universe is determined by the PQ mechanism: PQ symmetry breaking to make the axion possible and QCD instanton effects to give the axion mass.
Here we will concentrate on the production of a cosmic axion population, which primarily is expected to occur in the following ways.

\begin{itemize}

	\item Thermal axion production 
	
    Just as for other massive standard model particles, the mutual production and annihilation during reheating can result in a thermal relic population of QCD axions. 
    Concretely, QCD axions are produced from the standard model plasma typically by pion scattering, and decouple or freeze-out when the rate of the process $\pi+\pi\rightarrow \pi+a$ becomes lower than the Hubble rate. 
    The thermal axion abundance is determined by the decoupling temperature, with a roughly inverse relation, see e.g. Refs.~\cite{CosmosBook:Kolb&Turner, Berezhiani:1992rk}.
     
    ALPs are generically more weakly coupled to the standard model than the QCD axion, and their abundance is model dependent, with enough freedom to adjust.
	If ALPs arise in models with SUSY and extra dimensional compactifications (mainly string theory), a generic prediction states that the axion field can be coupled to a massive particle, $U$, with $m_U>m_a$. As $U$ decays, a population of relativistic ALPs is created~\cite{Acharya:2010zx, Cicoli:2012aq, Higaki:2012ar, Higaki:2013lra, Conlon:2013isa}. 
	This scenario is a kind of dark radiation with a rich phenomenology, but it is not the emphasis of our paper.

	\item Cold axion production 
	
	Various kinds of topological defects can be formed during the breaking of global symmetries~\cite{Kibble:1976sj}.
	In our case of a global $U(1)$  PQ symmetry, this gives rise to global axionic strings and domain walls (if the colour anomaly $\mathcal{C} > 1$, the domain walls are stable, otherwise unstable). 
	If PQ symmetry is broken after inflation, these topological defects can decay to produce a population of cold axions. But some other mechanism are typically needed to eliminate the domain wall problem~\cite{Hiramatsu:2012gg, Barr:2014vva}.
	
	In the scenario where PQ symmetry is broken before inflation, topological defects and their decay products are inflated away, and thus can be neglected. 
	
	Another method of cold axion production is vacuum realignment or misalignment\footnote{Here, ``vacuum realignment'' means the process that axion field relaxes to the potential minimum, and ``misalignment'' means the coherent initial displacement of the axion field}, which relies only on the basic properties of the axion \cite{PRESKILL1983127}.
	
	In an FLRW universe the equation of motion of the axion field is given by
	\begin{equation}
	\ddot{\phi}+3 H \dot{\phi} - {1 \over a^2} \bigtriangledown ^2 \phi + V_{eff}^{'}(\phi)~=0
	\label{eqn:axion_background_full}
	\end{equation}
	In the early Universe, we can reasonably assume the axion scalar field is uniformly distributed in space (i.e., homogeneous), then the spatial derivatives disappear, if we further take only the mass term in the effective potential for simplicity, we can get 
	\begin{equation}
	\ddot{\phi}+3H\dot{\phi}+m_a^2\phi=0
	\label{eqn:axion_background_simple}
	\end{equation}
	which is similar to the equation of a simple harmonic oscillator with time-dependent friction. 
	Furthermore, with $H(t_i)\gg m_a$, the initial conditions are well defined\footnote{At the time of PQ symmetry breaking the Hubble rate is much larger than the axion mass, so the field is overdamped and $\dot{\phi}=0$ initially.}:
	\begin{equation}
	\phi (t_i) = f_a \theta_{a,i} \, , 
	\quad \dot{\phi}(t_i) = 0 \, .
	\label{eqn:misalignment_initial_condition}
	\end{equation}
	Hence, the misalignment production of axions is determined by Eqs.~\eqref{eqn:axion_background_simple}, ~\eqref{eqn:misalignment_initial_condition}, while the coherent initial displacements of the axion field depends specifically on if PQ symmetry breaking occurs before or after inflation ~\cite{Marsh:2015xka}. 
	
\end{itemize}

\subsubsection{The condensation of axion dark matter }

After becoming massive, axions may intuitively evolve like ordinary dark matter, without special phenomena.
But thanks to the bosonic nature of axions and their very high phase space density, a number of novel condensation-derived effects have been suggested to occur~\cite{Sikivie:2009qn, Erken:2011vv, Erken:2011dz, Yang:2016odu, Oniga:2016ubb, Chakrabarty:2017fkd}, which are of use in the present context. 
The following parts of this section are mainly from Refs.~\cite{Sikivie:2009qn, Erken:2011vv, Erken:2011dz}, provided here to assist the reader in a better understanding overall.

We note that the underlying conditions for BEC formation are that: (i) the system comprise a large number of identical bosons, (ii)the bosons are conserved in number, (iii) the system are sufficiently degenerate and (iv) the system are in sufficient thermal equilibrium \cite{Chakrabarty:2017fkd}. CDM axions satisfy all the conditions, making the formation of a BEC a reasonable possibility.

However, before condensation can occur condition (iii) implies that axions firstly need to thermalize.  
One key quantity when discussing thermalization \& condensation is the relaxation rate, which roughly means the rate at which the system approaches a stable macroscopic state due to interactions between particles\cite{Sikivie:2009qn, Erken:2011vv, Erken:2011dz}, defined as
\begin{equation}
\Gamma \sim {\dot{\cal N} \over {\cal N}} 
\label{relax_def}
\end{equation}
where ${\cal N}$ is the occupation number of some key quantum states. 
To be clear, this relaxation rate is time-dependent.

As the relaxation rate reaches a value equal to or bigger than the Hubble rate, the system will approach thermal equilibrium and form a BEC. 
For convenience, we define another key quantity, the axion thermalization rate, as: 
\begin{equation}
    R_T = {\Gamma \over H}
\end{equation}
which is also time-dependent.

Before going on, let's introduce several other useful parameters:

(1)The axion particle number density in physical space is defined as
\begin{equation}
	n_a = {N_a \over V} = \int {d^3 p \over (2 \pi)^3} {n}_{a\_\vec p} 
	\label{particle_density}
\end{equation}
where $N_a$ is the total axion number in volume $V$, ${n}_{a\_\vec p}$ is the particle number density with momentum ${\vec p}$ \footnote{In \cite{Erken:2011dz}, it is expressed as ${\cal N}_{\vec p}$, but we think ${n}_{a\_\vec p}$ is more appropriate here.}.

(2) The velocity dispersion in the particle fluid is represented by $\delta v$.

(3) The correlation length of the particles is defined as $\ell_a \sim {1/\delta p} \sim {1/p_{\rm max}}$, where $p_{\rm max}$ is the maximum momentum.
	
We also emphasise here that there are two kinds of relaxation rate: the relaxation rate between axions themselves and the relaxation rate between axion and other particle species.

~\\
\textbf{Self-relaxation rate}

Based on the relation between the relaxation rate and the energy associated to the corresponding transition, there are two possible regimes: the ``particle kinetic regime'' and the ``condensed regime'', whose names are broadly indicative of the phase in which they are applicable.
We summarise the relevant results here, please check Ref.~\cite{Erken:2011dz} for their precise definition and further details. 

~\\
\begin{itemize}
\item{{\bf The particle kinetic regime}}

The estimate for the relaxation rate from $\lambda \phi^4$ self-interactions is:
\begin{equation}
\Gamma_{a\lambda\_k} \sim {\lambda^2 \over 64 \pi} {1 \over m_a^2}~n_a~\delta v~{\cal N}
\label{relax_kr_l}
\end{equation}
whilst the corresponding relaxation rate from gravitational interactions is:
\begin{equation}
\Gamma_{ag\_k} \sim {4 G^2 m_a^2 \over (\delta v)^4}~n_a~\delta v~{\cal N}
\label{relax_kr_g}
\end{equation}

~\\
\item{{\bf The condensed regime}}

The estimate for the relaxation rate from $\lambda \phi^4$ self-interactions is:
\begin{equation}
\Gamma_{a\lambda\_c} \sim {1 \over 4} \lambda n_a m_a^{-2}
\label{relax_cr_s}
\end{equation}
whilst the corresponding relaxation rate from gravitational interactions is:
\begin{equation}
\Gamma_{ag\_c} \sim 4 \pi G n_a m_a^2 \ell_a^2
\label{relax_cr_g}
\end{equation}

\end{itemize}

~\\
\textbf{Relaxation rates with other particle species}

Motivated by the question of whether other species, such as hot particles, photons and other cold species, can come into thermal contact with the cold axion fluid, and thus to explore more cosmic phenomenologies, it is important, especially for this paper's purpose, to construct the gravitational interaction relaxation rates of the cold axion fluid with these other species.

Generally, the relaxation rate due to interactions with other particle species $b$ can be represented as: 
\begin{equation}
	\Gamma_{b} \sim \frac{4 \pi G m_a n_a \ell_a \omega_b}{\Delta p_b} 
\end{equation}
where $\omega_b$ is the relativistic particle energy, which for non-relativistic low energy particles is approximately the particle mass $m_b$, and $\Delta p_b$ is the particle momentum dispersion. 
This formula can be simplified under different conditions as follows.

\begin{itemize}
\item{Hot particles}

The relaxation rate for relativistic particles, e.g. hot axions and non-degenerate neutrinos, interacting with the highly occupied low momentum axion modes is thus of order
\begin{equation}
\Gamma_r \sim 4 \pi G n_a m_a \ell_a
\label{relax_rate_of_hot_particles}
\end{equation}

\item{Cold particles}

For non-relativistic cold particles, such as hydrogen atoms, baryons/leptons and WIMPs (bosons or non-degenerate fermions), the relaxation rate is
\begin{equation}
\Gamma_B \sim 4 \pi G n_a m_a \ell_a {m_B \over \Delta p_B}
\label{relax_rate_of_cold_particles}
\end{equation}
where $\Delta p_B$ is their momentum dispersion.

\item{Photons}

The relaxation rate for photons is
\begin{equation}
\Gamma_\gamma \sim  4 \pi G n_a m_a \ell_a
\label{relax_rate_of_photons}
\end{equation}
which is the same as for hot particles, Eq.~(\ref{relax_rate_of_hot_particles}), in order of magnitude.

\end{itemize}
	
~\\
\textbf{Thermalization rates}

Axions are in thermal equilibrium if their relaxation rate $\Gamma_a$ is large compared to the Hubble expansion rate $H$. For convenience, we define axion self thermalization rate as: 
\begin{equation}
	R_{Ta} = {\Gamma_a \over H}
\end{equation}
where $H$ is the Hubble parameter.

We have noted that the formulae for the relaxation rate differ dependent on the regimes of the axion fluid (`particle kinetic' or `condensed'). Using Eq.~(\ref{relax_kr_l}) in the particle kinetic regime, we get the axion (and generic ALPs) self-thermalization rate due to the self-interaction $\lambda \phi^4$:
\begin{equation}
	R_{Ta\lambda\_k} = \frac{\Gamma_{a\lambda\_k}}{H} \propto \frac{ n_a }{ H m_a^2} \propto a(t)^{-3} t \propto t^{-{1 \over 2}}
	\label{sk_thermalize_rate_estimate}
\end{equation}
where $a(t)$ is the cosmic scale factor.

So there must exist a time $t_1$, before which the axions thermalize via this  $\lambda \phi^4$ self-interaction
\begin{equation}
	\Gamma_{a\lambda\_k}(t_1) \sim H(t_1)
	\label{thermalize_at_t1}
\end{equation}
and after which, the axions will fall out of thermal equilibrium.  

However, although around $t_1$ gravitational self-interactions are too weak to cause thermalization of cold axions, after $t_1$, the thermalization rate due to gravitational interactions is given by~\cite{Chakrabarty:2017fkd, Sikivie:2009qn}
\begin{equation}
	R_{Tag\_c} = \frac{\Gamma_{ag\_c}}{H} \sim \frac{4 \pi G n_a m_a^2 \ell_a^2}{H} \propto ~{a(t_1) \over a(t)} {t \over t_1} \propto a(t)
\end{equation}
This scales as $a(t)$, and is thus an increasing function of time. 

As a consequence, the $\lambda \phi^4$ interaction is only effective at thermalizing axions for a short period in the early Universe, whereas gravitational self-interactions can in contrast be effective at thermalizing axions over long time periods, and hence inducing them to form a BEC.

Once thermal equilibrium occurs, there can be cooling effects between axion and other particle species. That is to say, not only the interaction between the axions themselves, but also the gravitational interaction between the axions and any other particles can lead to thermal equilibrium. When the thermal equilibrium is achieved, there is a heating effect on the lower energy particles, and a cooling effect on other higher energy particles. 

The generalized thermalization rate between axion and another particle species $b$ is defined as: 
\begin{equation}
	R_{Tb} = {\Gamma_b \over H}\,
\end{equation}

Then the following condition is a criteria for thermal equilibrium, cooling and heating effects:
\begin{equation}
	R_{Tb} = \frac{\Gamma_{b}}{H} \sim \frac{4 \pi G m_a n_a \ell_a \omega_b}{H \Delta p_b } \gtrsim 1\,
	\label{axion_thermal_balance_condition}
\end{equation}

Here $b$ in particular can be Hydrogen(cold particles), then we have:
\begin{equation}
	R_{TH} = \frac{\Gamma_H}{H} \sim \frac{4 \pi G m_a n_a \ell_a m_H}{H \Delta p_H } \,
	\label{hydrogen cooling condition}
\end{equation}
where $m_H$ is the mass of Hydrogen atom, $\Delta p_H$ is the Hydrogen atom momentum dispersion. 
For the current scenario of an axion BEC cooling Hydrogen gas, we will refer to the thermalization rate as the cooling rate in the following paragraphs.

~\\
\textbf{Some comments}

Although there has been some controversy in the literature\cite{Saikawa:2012uk, Davidson:2013aba, Davidson:2014hfa} around the effect of interactions between axions and other particle species, beyond the original progenitors of the scenario it has for example been confirmed that an axion BEC can indeed form in Ref.~\cite{Guth:2014hsa}.
Some points remain unsettled at present, insofar as in Ref.~\cite{Guth:2014hsa} it is for example argued that the value of the resulting correlation length may be overly reliant on the criterion of homogeneity and hence reduced to be small, whilst in Ref.~\cite{Chakrabarty:2017fkd}, it is emphasized that a BEC can be inhomogeneous and nonetheless correlated over its whole extent, which can be arbitrarily large.

\section{Hydrogen cooling induced by axion condensation}

The phenomenon discussed in the previous section offers the possibility to explain the anomalous EDGES result, with condensed axion dark matter cooling the primordial hydrogen after it decouples from the CMB at $z \sim 200$.

This latter point is essential, as if axion cooling begins whilst the CMB and hydrogen are in thermal equilibrium, the effect on Eq.\eqref{T21} will be negligible.
Of course the onset of cooling must also occur prior to the cosmic dawn, and the effect in total must give the correct EDGES absorption magnitude.
As we will see in the following, and perhaps surprisingly, these various requirements can be simultaneously accommodated by an ALP which may also function as the QCD axion.
In practice the EDGES observation uniquely selects a small range for $m_a$, which is compatible with present-day axion phenomenology and can also conceivably be explored at the next generation of axion experiments.

Using the formulae from the previous section, our starting point is the baryon cooling rate (i.e. above mentioned thermalization rate) at the time of matter-radiation equality,
\begin{equation}
	\frac{\Gamma_H}{H} \sim \frac{4\pi G m_a n_a \ell_a m_H}{ H \Delta p_H} \simeq \frac{4\pi G m_a n_a}{H^2} \left(\frac{m_H}{3 T_H}\right)^{1/2}  
	\label{hydrogen_cooling_rate_raw}
\end{equation}
where $T_H$ is the temperature of Hydrogen gas,  and we have used the approximation $\rho_a \simeq m_a n_a$, which is sufficiently precise at low temperatures. 
Assuming that we are in the axion condensed phase, we have also identified the axion BEC correlation length $\ell_a \sim 1/H$ (this value is the extreme limit, before the BEC is formed, the correlation length is smaller, $\sim {1/p_{\rm max}}$ ~\cite{Sikivie:2009qn}).
By virtue of the Maxwell-Boltzmann distribution $\Delta p \simeq \sqrt{3 m_H T_H}$, and at this temperature we can identify $\omega \simeq m_H$.

Via the Friedmann equation $3H^2 \simeq 8 \pi G \rho_{tot}$ we also have 
\begin{equation}
\frac{\Gamma_H}{H} \sim \frac{4\pi G m_a n_a}{(8\pi G/3) \rho_{tot}} \left(\frac{m_H}{3 T_H}\right)^{1/2} \simeq \left(\frac{m_H}{3 T_H}\right)^{1/2} \frac{3\rho_a}{2\rho_{tot}}
\end{equation}

At the epoch of matter-radiation equality, $\rho_{tot} \simeq 2 \rho_{DM}$, then, we get
\begin{equation}
	\frac{\Gamma_H}{H} \bigg |_{t_{eq}} \sim \left(\frac{m_H}{3 T_H}\right)^{1/2} \frac{3\rho_a}{4\rho_{DM}}  \bigg |_{t_{eq}} = \left(\frac{3m_H}{16 T_{eq}}\right)^{1/2}\frac{\Omega_ah^2}{\Omega_{DM}h^2}
	\label{hydrogen cooling rate}
\end{equation}
neglecting the contributions of visible matter and dark energy. $\Omega_ah^2/\Omega_{DM}h^2$ is the fraction of the cooling-induced ALP density over the total dark matter relic density. 

As $m_H>>T_{eq}$ we evidently need a small $\left(\Omega_a/\Omega_{DM}\right)$ ratio to ensure cooling begins only when $z \in(200, 20)$. 
To be more precise we note that since $a\propto t^{2/3}$ during matter domination, we have
\begin{equation}
\frac{\Gamma_H}{H} \sim \frac{4\pi G m_a n_a}{H^2} \left(\frac{m_H}{3 T_H}\right)^{1/2} \propto \frac{4\pi G m_a N_a}{a^3 (\dot{a}/a)^2} \left(\frac{m_H}{3 T_H}\right)^{1/2} \propto {9\pi G m_a N_a} \left(\frac{m_H}{3 T_H}\right)^{1/2} \propto \left(\frac{1}{T_H}\right)^{1/2}
\end{equation}
where $N_a$ is the total axion number, and $a$ is the cosmic scale factor.
This means $\Gamma_H/H\propto 1/\sqrt{T_H}$, which implies that after matter-radiation equality,
\begin{equation}
	\frac{\Gamma_H}{H} = \frac{\Gamma_H}{H} \bigg |_{t_{eq}}\left(\frac{T_{eq}}{T_H}\right)^{1/2} = \left(\frac{3m_H}{16 T_{eq}}\right)^{1/2}\frac{\Omega_ah^2}{\Omega_{DM}h^2} \left(\frac{T_{eq}}{T_H}\right)^{1/2} = \left(\frac{3m_H}{16 T_{H}}\right)^{1/2}\frac{\Omega_ah^2}{\Omega_{DM}h^2} 
\end{equation}

To have the cooling effect, via Eq.\eqref{hydrogen cooling condition}, we set $\Gamma_H/H = 1$ to yield
\begin{eqnarray}
	\frac{\Omega_a h^2}{\Omega_{DM} h^2}=\left(\frac{16 T_{H}}{3m_H}\right)^{1/2}\,,\nonumber\\
	\Omega_a h^2 = \left(\frac{16 T_{H}}{3m_H}\right)^{1/2} \Omega_{DM} h^2 
	\label{CDM_relic_density_relation}
\end{eqnarray}
Since $T_{eq} \sim 0.75$ eV $\simeq 8.7 \times 10^3$ K, and we require axion-induced cooling to occur between $T_H^{z=200} \sim 475$ K and $T_H^{z=20} \sim 10$ K, we can firstly establish the requirement
\begin{equation}
	\frac{\Omega_a h^2} {\Omega_{DM}h^2} \in (0.22,1.5)\times10^{-5}
	\label{CDM_relic_density}
\end{equation}

It is important to note that once condensation occurs, we will have two distinct populations of cold axions; those that are in the condensed state, and a remnant thermal population.
Hydrogen can in principle interact with both, however there exists a key distinction: scattering from the cold thermal axions will simply raise their temperature, whilst scattering condensed axions will typically liberate them from the BEC, given the energies involved, and into the thermal population.

However, in Ref.~\cite{Davidson:2014hfa} the rate at which the BEC occupation number can change via scattering with external particles is calculated, finding that the latter number-changing process should be vanishingly rare.
To be careful though, this only means that the BEC cannot be excited directly by the hydrogen, and does not exclude that the BEC cannot rethermalize by itself.
As such the total BEC axion occupation number density can in principle change.

Since the energy lost from the hydrogen must be transferred to the thermal axions, energy conservation requires
\begin{equation}
	\rho_H(T_{Hi}) + \rho_{ac}(T_{ai}) + \rho_{at}(T_{ai})  = \rho_H(T_f) + \rho_{ac}(T_f)  + \rho_{at}(T_f)
	\label{energy_conservation_total}
\end{equation}
Here, $T_{Hi}$, $T_{ai}$ are the initial (pre-cooling) hydrogen, BEC axion temperatures, respectively. $T_f$ is the final (post-cooling) temperature of hydrogen and BEC axion. In fact, $T_{Hi}$ is just equal to $T_H$ of the $\Lambda$CDM before BEC axion cooling, here for comparing with the $T_f$, we use $T_{Hi}$, especially in the following cooling related equations, and always use $T_H$ as the usual hydrogen temperature of $\Lambda$CDM. $\rho_H(T_{Hi})$, $\rho_{ac}(T_{ai})$, $\rho_{at}(T_{ai})$ are the initial (pre-cooling) hydrogen, BEC axion, and thermal axion energy densities, respectively. Meanwhile $\rho_H(T_f)$, $\rho_{ac}(T_f)$, $\rho_{at}(T_f)$ are the final (post-cooling) hydrogen, BEC axion, and thermal axion energy densities, respectively.

Because $T_{ai}$ is very low, $\rho_a(T_{ai})$ is small enough to be ignored. So that the conservation equation becomes
\begin{equation}
	\rho_H(T_{Hi}) \simeq \rho_H(T_f) + \rho_{at}(T_f) - \rho_{at(\text{rest mass})}
	\label{energy_conservation}
\end{equation}
where, $\rho_{at(\text{rest mass})} = \rho_{ac}(T_{ai}) - \rho_{ac}(T_f)$ means the rest mass energy density of the axions entering the thermal axion population from the BEC axion population via rethermalization
\footnote{This is strictly a slightly more precise energy conservation equation than that used in Ref.~\cite{Houston:2018vrf}, although the effect of the additional term on our final result is negligible thanks to the 4th order root extraction.}, \footnote{For our parameter range of interest, both photon cooling by axions and thermal axion heating by photons is strongly suppressed, as there is no large $\sqrt{m_H/(3T_H)}$ factor in the corresponding photon cooling rate, see Eqs.\eqref{relax_rate_of_photons}, \eqref{relax_rate_of_cold_particles}, \eqref{hydrogen_cooling_rate_raw}. 
We also note the principal constraint in the axion-induced cooling ${}^7$Li scenario was a large resulting $N_{\mathrm{eff}}$ at recombination.
For us this is not a cause for concern as we are operating at a much later epoch, and the thermal axions excited will be non-relativistic.}.

In the case of cold hydrogen, to the lowest order, we have
\begin{equation}
\rho_H(T_H) \simeq n_H(m_H + 3T_H/2)
\label{energy_density_hydrogen}
\end{equation} 
where $n_H$ is the Hydrogen number density. 
Since hydrogen comprises the vast majority of baryonic matter at this epoch we can use the baryon-to-photon ratio to estimate
\begin{equation}
n_H \simeq 6\times 10^{-10}\,n_\gamma
\label{number_density_hydrogen}
\end{equation} 
where $n_\gamma$ is the photon number density
\begin{equation}
n_\gamma = 2\zeta(3)T_\gamma^3/\pi^2
\label{number_density_photon}
\end{equation}

For simplicity, we can assume
\begin{eqnarray}
T_\gamma \simeq {{T_\gamma^{z=200} \times \frac{z + 1}{200+1}} } \,,\nonumber\\
T_H \simeq T_H^{z=200} \times \left(\frac{z + 1}{200+1}\right)^2 
\label{temperature_hydrogen_photon}
\end{eqnarray}
Where the formula for $T_\gamma$ is accurate enough, but the formula for $T_H$ is not so accurate. 
We provide this formula primarily to illustrate how the final result is computed, in practice we use RECFAST to get a more accurate relation in the following sections.

Inserting a Maxwell-Boltzmann distribution for the thermal axion population we have:
\begin{itemize}
\item The energy density of thermal axions
\begin{equation}
	\rho_{at}(T)=\frac{T^4}{2\pi^2}\int_{0}^\infty\frac{\xi^2\sqrt{\xi^2+(m_a/T)^2}}{\exp\left(\sqrt{\xi^2+(m_a/T)^2}\right)- 1}d\xi
	\label{energy_density_axion_thermalized}
\end{equation}

\item The number density of thermal axions
\begin{equation}	
	n_{at}(T) = \frac{T^3}{2\pi^2}\int_{0}^\infty\frac{\xi^2}{\exp\left(\sqrt{\xi^2+(m_a/T)^2}\right)- 1}d\xi
	\label{number_density_axion_thermalized}
\end{equation}

\item The rest mass energy density of the rethermalized axions
\begin{equation}	
	\rho_{at(\text{rest mass})} = (n_{at}(T_f)-n_{at}(T_{ai})) m_a \simeq n_{at}(T_f) m_a =\frac{T_f^3 m_a}{2\pi^2}\int_{0}^\infty\frac{\xi^2}{\exp\left(\sqrt{\xi^2+(m_a/T_f)^2}\right)- 1}d\xi 
	\label{rest_mass_density_axion_thermalized}
\end{equation}
\end{itemize}

Putting all the Eqs.\eqref{energy_density_hydrogen}, \eqref{energy_density_axion_thermalized}, \eqref{rest_mass_density_axion_thermalized}, into eq.\eqref{energy_conservation}, we get a complicated expression for the final energy conservation equation, with three unknown variables $m_a$, $T_{Hi}$ and $T_f$. 
If given $T_{Hi}$ and $m_a$, we can solve this equation numerically for $T_f$ and hence the cooling ratio $T_f/T_{Hi}$.

By Eq.\eqref{temperature_hydrogen_photon} and assuming the change in $z$ is negligible during the cooling process, we find the cooled Hydrogen temperature at redshift $z$ is
\begin{equation}
	T_{Hc} \simeq T_{f} \left(\frac{z+1}{z_{c}+1}\right)^2 =  T_{Hi} \left(\frac{z+1}{z_{c}+1}\right)^2 \left(\frac{T_f}{T_{Hi}}\right) = T_H \left(\frac{T_f}{T_{Hi}}\right)
\end{equation}
where $z_{c}$ is the redshift at which cooling begins, $z < z_{c}$, and $T_H$ takes its usual $\Lambda$CDM value
\footnote{This assumption that $z$ is unchanged during the cooling process, or in other words that the cooling mechanism between hydrogen and axions is relatively efficient, is valid because the characteristic timescale of these processes in the condensed phase is faster than that of ordinary gravitational evolution, as demonstrated in detail in \cite{Erken:2011dz}.}.

Since $T_s = T_H$ at this epoch, we then find the cooled spin temperature is
\begin{equation}
	T_{sc} \simeq T_H \left(\frac{T_f}{T_{Hi}}\right) 
\end{equation}
Putting this result into Eq.\eqref{T21}, we arrive at
\begin{equation}
	T_{21}=35 \,\mathrm{mK}\left(1-\frac{T_{Hi}}{T_f}\frac{T_\gamma}{T_H}\right)\sqrt{\frac{1+z}{18}}
	\label{Cooled_T21}
\end{equation}
where $T_\gamma$ and $T_H$ take their usual $\Lambda$CDM values.

From the EDGES result, $T_{21}$ is limited in a range:
\begin{equation}
T_{21} ^{z \sim 17} \in (-0.94, -0.34) \,\text{K}
\label{T21_Limitation}
\end{equation}
This limit condition will give some important constraints on the axion parameter space, which will be illustrated in the following section. 

For a brief overview of the cooling process, we offer the following schematic graph:
\begin{figure}[h!]
	\centering
	\includegraphics[width=1.0\linewidth]{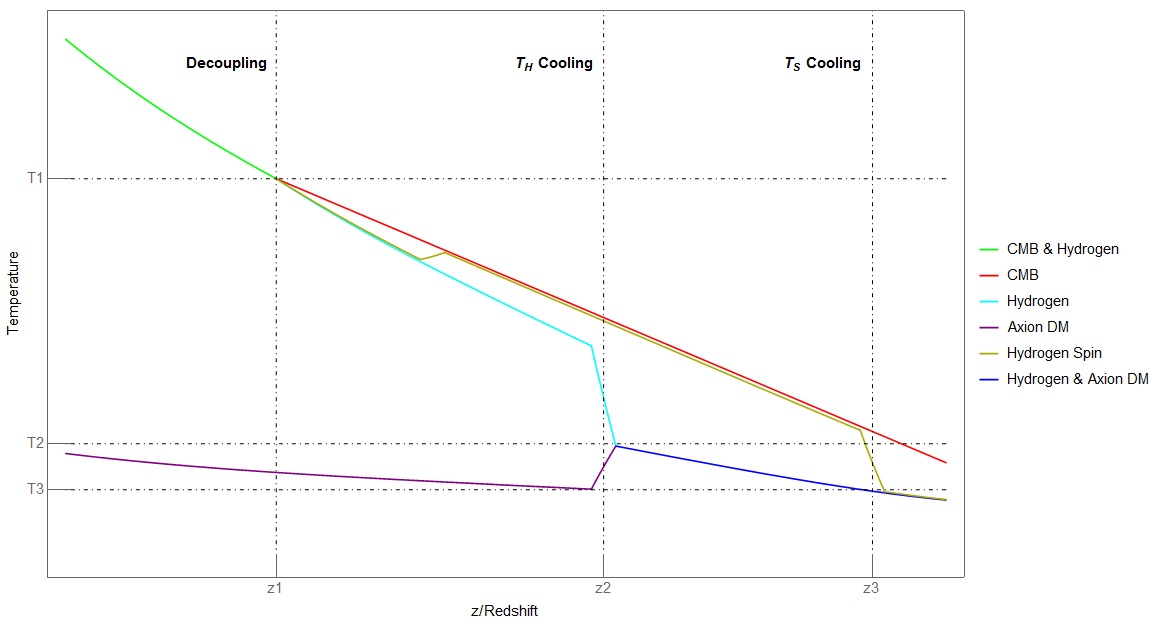}
	\caption{Schematic overview of the axion-hydrogen cooling process. Here z1 is the redshift when hydrogen and the CMB decoupled at the temperature T1; z2 is the redshift when hydrogen was cooled by the axion dark matter to the temperature T2; and z3 is the redshift when the hydrogen spin temperature was coupled to the hydrogen temperature via the Wouthuysen-Field effect and so reduced to the temperature T3. }
	\label{total_cooling_process}
\end{figure}
 
\section{Numerical computation and experimental constraints}

\begin{figure}[h!]
	\centering
	\includegraphics[width=0.8\linewidth]{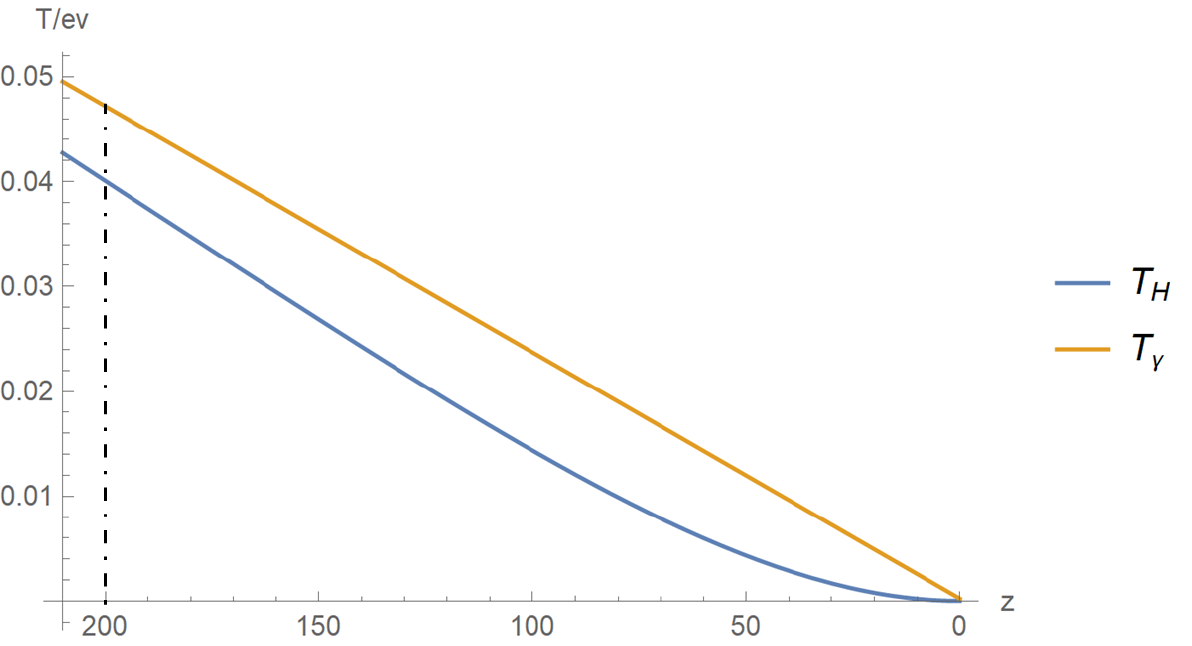}
	\caption{The temperature evolution of Hydrogen gas and the CMB.}
	\label{Fitting Temperature}
\end{figure}

Based on the previous derivations and formulae, we can then now calculate our desired result.
In practice additional care is needed since the basic redshift relations do not accurately capture the evolution of $T_H$ in this region, so we use RECFAST to compute $T_H$ and $T_\gamma$ \cite{Seager:1999bc}.
Numerically fitting we find
\begin{align}
	T_H&= {{0.267628 + 0.0034228 z + 0.0241129 z^2 -0.0000872526 z^3 + 1.30977 \times 10^{-7} z^4-3.33753 \times 10^{-11} z^5} \over 11605}\,, \nonumber\\
	T_\gamma&= {{2.725 + 2.725 z - 1.05\times 10^{-18} z^2} \over 11605}
\label{Hydrogen_temperature_in_LCDM}
\end{align}
as shown in Figure \ref{Fitting Temperature}.

We can find CMB temperature is almost linear with redshift $z$, while hydrogen gas temperature is more complicated, and in fact the decoupling of the CMB and hydrogen gas begins earlier than $z = 200$. The new accurate fitting relation of $T_H$ and $z$ is the basic reason of the numerical differences between the present paper and our previous short paper ~\cite{Houston:2018vrf}. 

The resulting dependence in Eq.\eqref{Cooled_T21} is however nonetheless correct, and so we can use Eq.\eqref{energy_conservation} to find the resulting 21 cm absorption feature.

\subsection{The mass and relic density constraint}

The energy conservation equation Eq.\eqref{energy_conservation} has three unknown variables $m_a$, $T_{Hi}$, $T_f$. Here $T_{Hi}$ is determined by $z_c$. For practical computation we construct an array of values for $z_c$ and $x = m_a/T_f$ respectively, then for every pair of ($z_c, x$), solve for the value of $m_a$, and hence $T_f = m_a/x$. 
So, for a certain value $m_a$, there are many different ($z_c, x$) or ($T_{Hi}, T_f$) pair values.
Equivalently, by equation Eq.\eqref{CDM_relic_density_relation}, $\Omega_a h^2$ is determined by $T_{Hi}$, so we can also say for a certain value $m_a$ there are many different possible ($\Omega_a h^2, T_f$) values.

\begin{figure}[h!]
	\centering
	\includegraphics[width=0.8\linewidth]{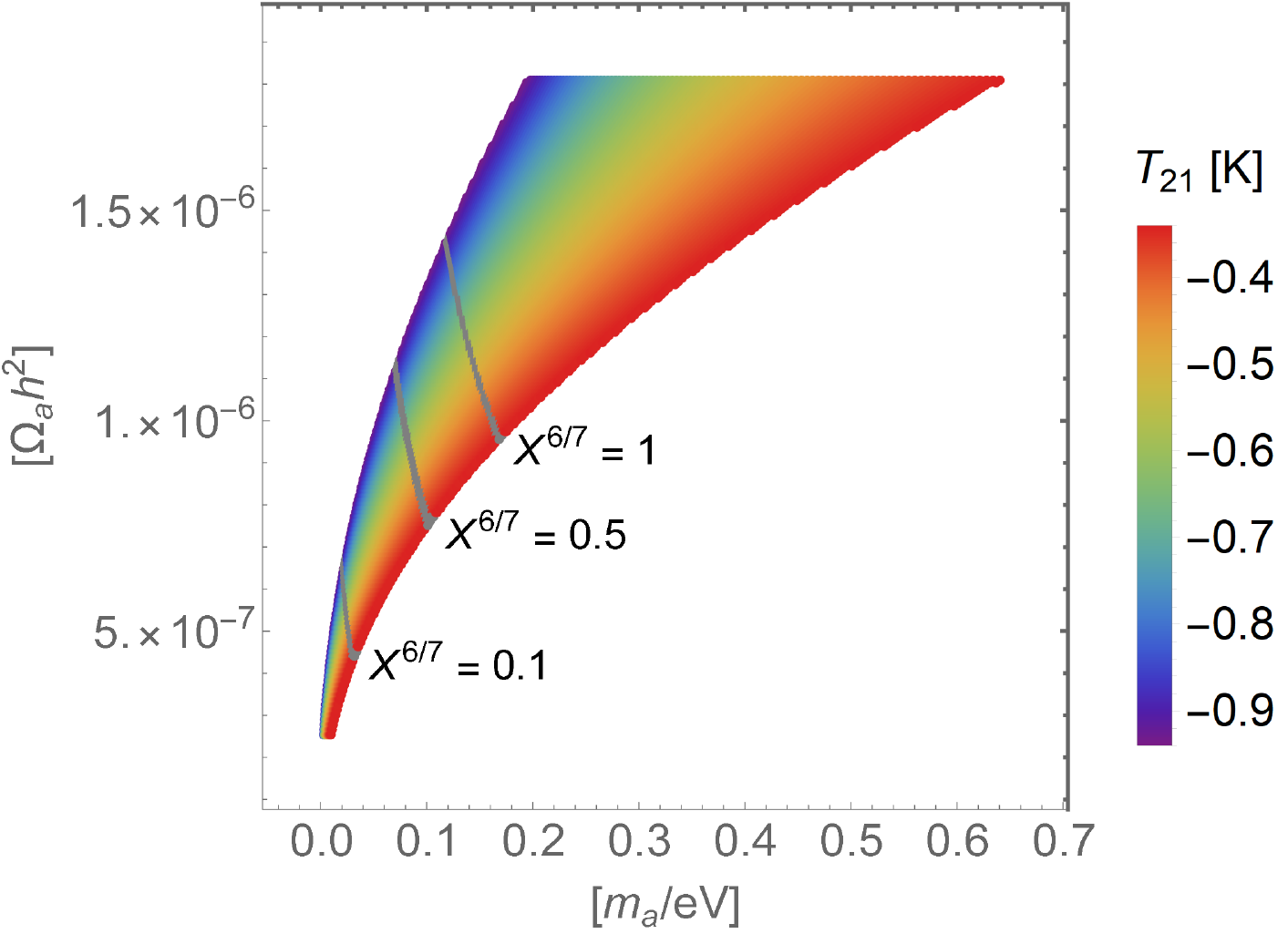}
	\caption{The ALP $(m_a,\Omega_ah^2)$ parameter space satisfying Eq.\eqref{CDM_relic_density}, colour-coded with the resulting 21cm brightness temperature at $z=17$.
	Comparison with the best-fit EDGES result suggests a $m_a\in(6,400)$ meV range of compatibility.
	Since the QCD axion fixes the relationship between these quantities in terms of the dark matter density parameter $X$ appearing in Eq.\eqref{QCD_axion_relic_density}, we overlay lines of fixed $X$ to show dependence on this quantity.
	}
	\label{relic_density_constraint}
\end{figure}

From Eq.\eqref{Cooled_T21}, we can find $T_{21}$ in terms of two independent variables, $m_a$ and $\Omega_a h^2$. 
From the ranges given by \eqref{T21_Limitation} and \eqref{CDM_relic_density}, we can then get a constraint on $m_a$ and $\Omega_a h^2$ as given in Fig.~\ref{relic_density_constraint}. 
By virtue of the EDGES best-fit result, where each value gives $T_{21} \simeq -0.52$ K at $z \sim 17$, we will favour an ALP with mass $m_a \in (6, 400)$ meV.
If we relax $T_{21}\in(-0.94, -0.34)$ K at $z \sim 17$\footnote{Here we remind the reader that $T_{21} \simeq -0.21$ K is the standard $\Lambda$CDM result, which we reach in the limit of this mechanism being inoperative.}, the 99\% confidence limits presented in Eq.\eqref{T21}, ALP mass will get a bigger range of $m_a \in (2.7, 640)$ meV.

Since in the generic ALP case the relationship between $\Omega_a h^2$ and $m_a$ is unfixed, we cannot directly connect them to coupling constraints and thus standard axion phenomenology. However, for the QCD axion the corresponding $f_a$ is given via
\begin{equation}
	\Omega_a h^2 = 0.15 X \left(\frac{f_a}{10^{12}\,\mathrm{GeV}}\right)^{7/6}
	\label{QCD_axion_relic_density}
\end{equation}
where $X$ is scenario-dependent unknown parameter. 
For PQ symmetry breaking prior to inflation typically $X \simeq \sin^2\theta_{mis}/2$, whilst for PQ symmetry breaking after inflation typically $X \in \left(2,10\right)$ depending on the relative contributions of topological defect decays and vacuum misalignment \cite{Erken:2011dz}. Combing these two scenarios, generally we have $X \in \left(0,10\right)$.
By Eq.\eqref{CDM_relic_density}, this implies that 
\begin{equation}
f_a\in\left(1.2,6.1\right)\times X^{-6/7}\times 10^7 \,\mathrm{GeV} 
\end{equation}
which is in turn related through chiral perturbation theory to $m_a$ via
\begin{equation}
	m_a \simeq 6\,\mathrm{eV}\left(\frac{10^6\,\mathrm{GeV}}{f_a}\right)
	\label{axion_mass_range}
\end{equation}
yielding 
\begin{equation}
	m_a \simeq 1.18\times 10^{-6} \,\mathrm{eV}\left(\frac{\Omega_a h^2}{X}\right)^{-6/7}
	\label{axion_mass_and_relic_density_relation}
\end{equation}
where care is required in that $m_a$ is now not freely varied in this instance; each value is associated to a specific $\Omega_a h^2$, and thus the specific $z_c$ and $T_{Hi}$ at which cooling begins.
Additionally by Eq.\eqref{CDM_relic_density}, numerically, we get
\begin{equation}
	m_a \in \left(0.1,0.5\right) \times X^{6/7}\, \mathrm{eV}
\end{equation}

Taking care to accommodate this, we arrive at a one-to-one mapping between $m_a$ and $T_{21}$.
We also note for completeness that in this mass range we can expect both hot and cold axion dark matter, due, for example, to thermal production and vacuum misalignment respectively.

Since $X\in(2,10)$ for post-inflationary PQ symmetry breaking, the minimum value for this quantity is realized for pre-inflationary symmetry breaking, in which case we have $X^{6/7} \sim 0.5$ in the absence of fine-tuning, assuming the initial misalignment angle is randomly drawn from a uniform distribution on $[-\pi,\pi]$, giving $\langle\theta_{\mathrm{mis}}^2\rangle=\pi^2/3$.

Varying $X$ we will find a preferred natural range of $m_a\in(6,400)$ meV for the QCD axion by virtue of the EDGES best-fit result(where each value gives $T_{21} \simeq -0.52$ K at $z \sim 17$ as mentioned before), this is same with the ALP situation. 

If we fix $X^{6/7}=1$ as a benchmark case and relax $T_{21}\in(-0.94, -0.34)$ K at $z \sim 17$, working backwards, then in this case imply $m_a\in(120,170)$ meV, with the best fit value corresponding to $m_a \simeq 150$ meV.
For more $X^{6/7}$ benchmark values, see the following table.
\begin{table}[h]
\begin{tabular}{|c|c|c|c|}
\hline
$X^{6/7}$ benchmark & min mass (/meV)  & best fit mass (/meV)  & max mass (/meV) \\
\hline
$1$ & $120$ & $150$ & $170$ \\
\hline
$0.5$ & $73$ & $90$ & $100$ \\
\hline
$0.1$ & $23$ & $28$ & $33$ \\
\hline
\end{tabular}
\caption{The QCD axion mass range for several $X^{6/7}$ benchmark values.}
\label{tab:QCD_axion_mass}
\end{table}

Considering all these ALP and QCD axion possibilities, we show the parameter space in Fig.~\ref{relic_density_constraint}, where the QCD axion is represented via lines of constant $X^{6/7}$.

\subsection{The mass and coupling constraint}

\begin{figure}[h!]
	\centering
	\includegraphics[width=1\linewidth]{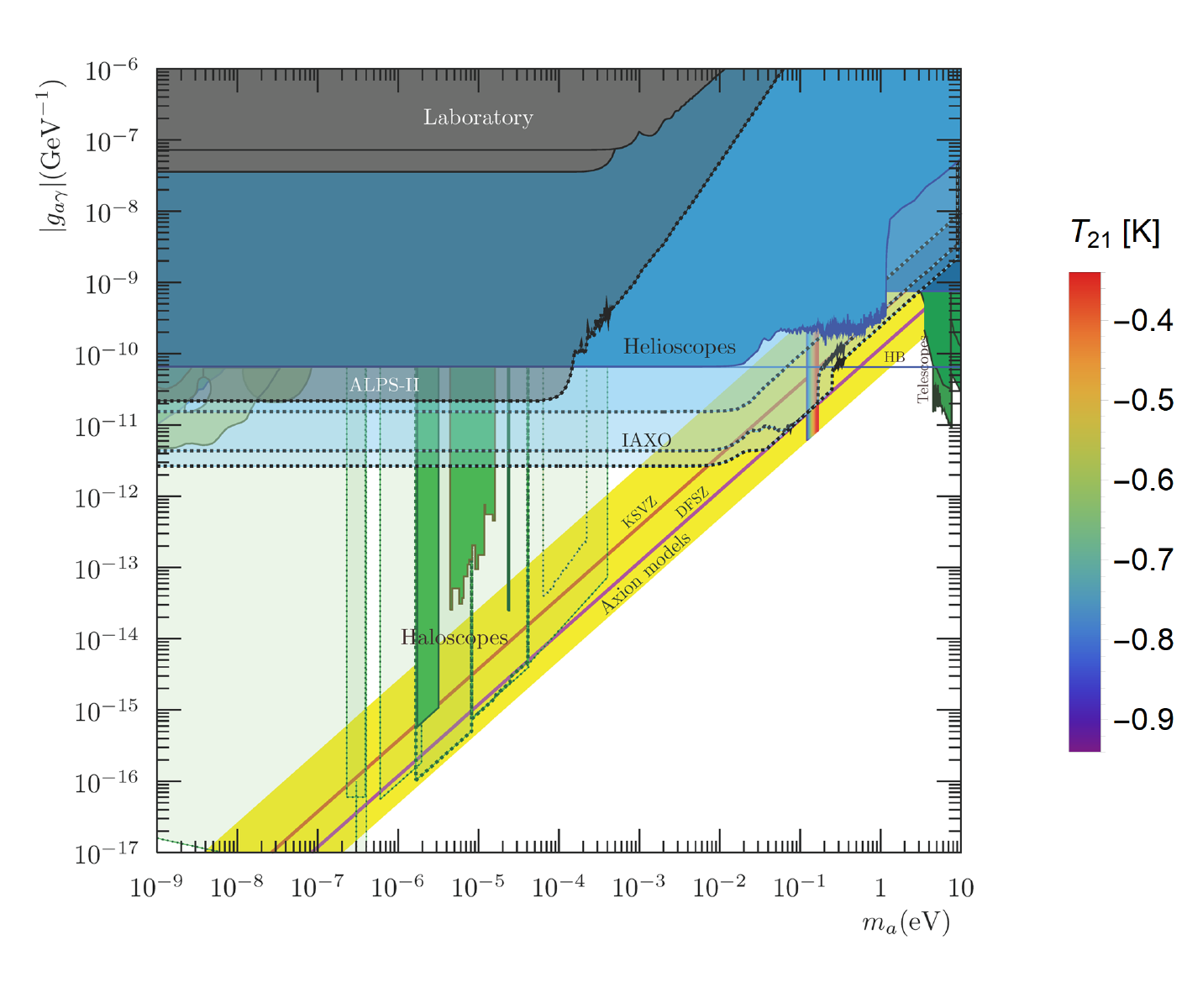}
  	\caption{The region of the axion parameter space relevant for our purposes, reproduced from \cite{Irastorza:2018dyq}, with the 21cm brightness temperature at $z \sim 17$ overlaid from axion-induced cooling processes in the benchmark case of $X=1$.
	The yellow band denotes QCD axion models with varying electromagnetic/colour anomaly coefficients, whilst the black curves indicate forecast sensitivities for the proposed IAXO experiment.
	The best fit $m_a$ value preferred by the EDGES observations in this case is 150 meV.
	}
	\label{coupling_constraint}
\end{figure}

Since the generic ALP case does not immediately translate to ordinary axion coupling constant constraints, we can specialize to the QCD axion to gain some phenomenological insight and delineate the parameter values implied by the EDGES observation in this scenario, along with the various experimental and observational constraints which may apply.
In Fig.~\ref{coupling_constraint} we reproduce constraints on the axion parameter space in our region of interest from \cite{Irastorza:2018dyq} colour coded with the resulting value of $T_{21}$ at $z=17$ for the benchmark case of $X=1$ (That is $m_a\in(120,170)$ meV).
As is evident, the EDGES observations can be straightforwardly accommodated within the ordinary QCD axion band.

It is of course important to note that the full possible mass range favoured by these results is strongly disfavoured for DFSZ type axions due to stellar energy-loss constraints \cite{Zhitnitsky:1980tq, Dine:1981rt, Graham:2015ouw}. 
As such we are implicitly considering KSVZ type axions \cite{Shifman:1979if, Kim:1979if}, although the ratio $E/N$ of the electromagnetic to colour anomaly is however allowed to vary within the usual range to accommodate variant models of the QCD axion \cite{DiLuzio:2016sbl, DiLuzio:2017pfr}.

Strictly speaking even then there is some tension between our preferred mass range and the observed burst duration of SN1987A, which favours $f_a\gtrsim 4\times 10^8$ GeV for standard QCD axions \cite{Raffelt:1996wa}.
This arises from an inference of the SN1987A cooling timescale, and thus energy loss to axions, from the time interval between the first and last neutrino observation.
However, given that these limits are derived from a single observation, and not to mention our limited knowledge regarding about axion emission in this extreme environment (the resulting exclusion being `fraught with uncertainties' in the words of Ref.~\cite{Raffelt:1996wa}), we can follow the example of others (e.g. Ref.~\cite{Archidiacono:2013cha}) and exercise a measure of caution in applying this constraint.

Furthermore, in Ref.~\cite{Giannotti:2017hny}, the top two panels of Figure 2 suggest that the mass range in question can still be compatible with DFSZ axions in a region favored by stars, and furthermore in fact seems to be particularly interesting for DFSZ-II models.
We also notice that in a recent preprint ~\cite{Lee:2018lcj} the author argues that those bounds are actually overestimated by an order of magnitude and recalculates to find weaker constraints, which would then marginally permit e.g. DFSZ axions in our 21 cm scenario.
So-called `astrophobic' axion models are also noteworthy here, where $\mathcal{O}$(100) meV axion masses are allowed at the cost of introducing some flavour-violating couplings \cite{DiLuzio:2017ogq, Hindmarsh:1997ac}.

In addition, we can also recapitulate at this point that ultimately the axion cooling mechanism employed here is gravitationally mediated, and so could be achieved with no Standard Model couplings whatsoever, and thus no issues in this regard.
By extension, the use of the QCD axion is in this context non-essential, and our primary results for generic axion-like-particles can still apply regardless.

We can also note from Ref.~\cite{Archidiacono:2013cha} that although our mass range of interest evades hot dark matter constraints at present, future large scale surveys such as the EUCLID mission are in conjunction with Planck CMB data projected to probe $m_a\gtrsim150$ meV for the QCD axion at high significance, allowing this scenario to be definitively tested in the near future \cite{Archidiacono:2015mda}.
More specifically, EUCLID will probe at the expansion of the universe and large scale structure, via galaxy observations out to $z \sim 2$.
Axions can be produced thermally if their couplings are not extremely suppressed, and so act as hot dark matter, which has a very different phenomenology compared to cold dark matter (although we note that cold and hot dark matter populations can exist simultaneously, so this is not necessarily an impediment to the cold dark matter condensation-based phenomenon we explore herein).
Once the mass of QCD axion is above 150 meV, it will have couplings strong enough that it can be copiously produced via thermal processes during the QCD phase transition \cite{Berezhiani:1992rk}, resulting in too much hot dark matter, which alters structure formation in a way detectable by EUCLID.
This being the case, one may consider the ALP cooling scenario to be more flexible than that of the QCD axion, although we note at present the latter can still evade bounds due to excess hot dark mattter.

Inspired from Ref.~\cite{Sikivie:2018tml}, we also note that our mechanism may have a damping effect on Baryon Acoustic Oscillations (BAO). 
Although it is ultimately argued there that the net effect on BAO should be consistent with observations, it may be worthwhile to more deeply explore the consequences of this scenario for this and other cosmological observables. 
As such BAO related experiments, such as eBOSS \cite{eBOSS01, eBOSS02}, may also be of relevance to this scenario.

\section{Summary}

The EDGES collaboration have recently presented an anomalously strong 21cm absorption profile, which could be the result of dark matter interactions around the time of the cosmic dawn.
Despite a flurry of interest there is as of yet no clear consensus on the provenance of this effect, and whether it is indeed a signature of dark matter at all. 
However, these results nonetheless provide an exciting first window into a previously unexplored epoch.

We have in this paper explored the potential of condensed-phase axion dark matter to explain these anomalous observations via a reduction of the hydrogen spin temperature during this epoch.
By fixing the axion CDM relic density so that cooling begins within the appropriate epoch, we find cooling effects that are both capable of explaining the EDGES observations and compatible with present day axion phenomenology.
More specifically, we find that the EDGES best-fit result of $T_{21} \simeq -0.52$ K and the requirement that hydrogen cooling occur within the range $z\in (200,20)$ are consistent with the cooling induced by an axion-like-particle of mass $m_a\in(6,400)$ meV. 
Future experiments and large scale surveys, particularly the International Axion Observatory (IAXO) and EUCLID, should have the capability to directly test this scenario.

As however the underlying cooling mechanism relies only upon gravitational couplings, it is not limited strictly to the context of models of the QCD axion.
As such it may also be arranged to occur in the primary scenario of axion-like-particles with no Standard Model couplings whatsoever, which could then evade bounds from stellar cooling and supernova observations.

\begin{acknowledgments}

This research was supported by Scientific Research Starting Project YJ202011 for Advanced Imported Talents of Wuyi University, by a CAS President's International Fellowship, by the Projects 11875062, 11947302 and 11875148 supported by the National Natural Science Foundation of China, and by the Key Research Program of Frontier Science, CAS. 
The numerical results described in this paper have been obtained via the HPC Cluster of ITP-CAS, Beijing, China.

\end{acknowledgments}

\end{document}